\documentclass[twocolumn,showpacs,preprintnumbers,amsmath,amssymb]{revtex4}

\usepackage{graphicx}
\usepackage{dcolumn}
\usepackage{bm}

\begin{document}



\title{Novel precursors for boron nanotubes: the competition of two-center 
and three-center bonding}

\author{Hui Tang} \affiliation{Department of Applied Physics,
  Yale University, New Haven, CT 06520}

\author{Sohrab Ismail-Beigi} \affiliation{Department of Applied Physics,
  Yale University, New Haven, CT 06520}

\date{\today}

\begin{abstract}
We present a new class of boron sheets, composed of triangular and hexagonal 
motifs, that are more stable than structures considered to date and thus 
are more likely to be the precusors of boron nanotubes.  We describe 
a simple and clear picture of electronic bonding in boron sheets and 
highlight the importance of three-center bonding and its competition with 
two-center bonding, which can also explain the stability of recently 
discovered boron fullerenes. Our findings call for reconsideration of the 
literature on boron sheets, nanotubes and clusters.

\end{abstract}



\maketitle

All boron nanotubes (BNT), regardless of diameter or chirality, are
predicted to be metallic and have large densities of states (DOS) at their
Fermi energies (E$_F$) \cite{Boustani97europhys}. In contrast, carbon nanotubes (CNT) can be
semiconductors or metals with small DOS at their E$_F$. Metallic CNT are used
widely to study one-dimensional (1D)  
electronics \cite{Tans97,Bock97} and are superconducting at low
temperatures \cite{Kociak01, Takesue06}.  Due to the larger DOS, BNT should be
better metallic systems for 1D electronics and may have higher
superconducting temperatures than CNT.

Recent experiments have fabricated boron nanotubular structures both as
small clusters \cite{Kiran05} and  long, 1D geometries
\cite{Ciuparu04}. Understanding the properties of
BNT is crucial for realizing their applications.  For CNT, it has been
 fruitful to study two-dimensional (2D) graphene: \textit{e.g.}, many 
properties of CNT are derived directly from graphene \cite{
Hamada92, Jishi93}. For boron, however, no 2D planar structure
exists in its crystals which are built from B$_{12}$ icosahedra
\cite{Muett67}. Researchers have proposed several 2D boron
sheets (BS). The hexagonal graphitic BS was found to be unstable
\cite{Boustani99jcp, Evans05}.  Based on extensive theoretical studies of boron clusters \cite{Boustani97surfsci, Boustani97prb, Boustani99jcp, Boustani99, Boustani03}, an Aufbau principle was proposed whereby the most stable structures should be composed of buckled triangular motifs \cite{Boustani97prb} . Experiments on small clusters of 10-15 atoms support this view \cite{Zhai03}.  A recent study of many possible sheet structures found, again, the buckled triangular arrangement to be most favorable \cite{Lau07}.
Hence, 2D triangular BS have been studied and used 
to construct BNT \cite{Evans05, Kunstmann06, Cabria06}.  

In this Letter, we present a class of boron sheets that are more stable
than the currently accepted ones.  We describe their structures, energetics,
electronic states, and provide a clear picture of the nature of their 
bonding that clarifies their stability. We also show that clusters with these structures are competitive with or more favorable than those considered to date.  Our findings have important consequences for understanding and interpreting the properties of these systems.  
For example, the unusual stability of B$_{80}$ fullerenes \cite{Szwacki07} can be explained by our bonding picture.  Hence, in our view, it is necessary to reconsider previous work in this general field.  

\begin{figure}[b!]
\includegraphics[width=2.8in]{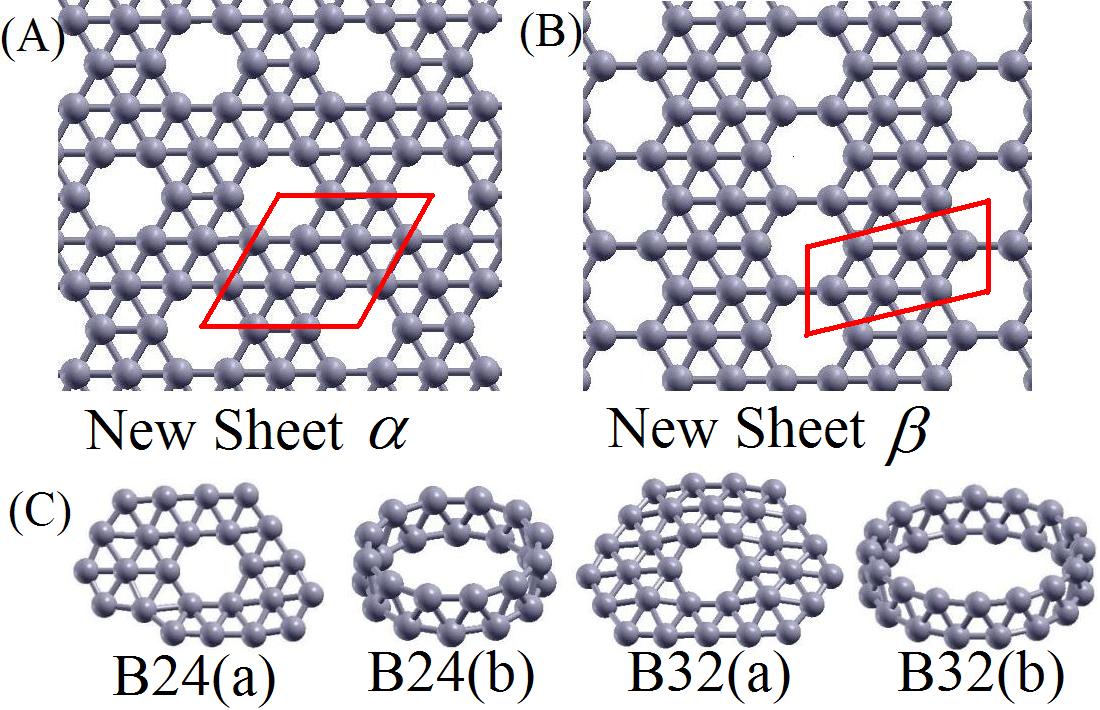}
\caption{\label{fig:AB}(A, B) Two examples of our  BS (top view).
Red solid lines show the unit cells.
(C) Four boron clusters: B$_{24}(a)$ and B$_{32}(a)$
are clusters with hexagonal holes; 
B$_{24}(b)$ and B$_{32}(b)$ are the double-ring clusters from refs. \cite{Boustani03, Boustani99}.  Gray balls are boron atoms, and gray ``bonds'' are drawn between nearest neighbors.
}
\end{figure}

We use Density Functional Theory \cite{Kohn65, Hohenberg64}
within the {\it ab initio} supercell planewave pseudopotential total energy
approach \cite{Payne92}. Calculations are done by PARATEC
\cite{para}. We use both the local density approximation (LDA)  
\cite{Kohn65, Perdew81} and the generalized gradient approximation (GGA)
\cite{Perdew92} for exchange and correlation.  Most results below employ 
the LDA and key results are checked by the GGA.  The LDA and
GGA yield same qualitative results with minor quantitative
variances. The planewave basis has a 32 Ryd cutoff energy.  K-point samplings for each case converge total energies to better than 1 meV/atom.  Norm-conserving pseudopotentials have cutoff radii
$r_c^s$=$1.7$  and $r_c^p$=$2.1 $ a.u..
The BS are extended in x-y directions while supercells have
periodic copies along z where a separation of 7.4 \AA \ is sufficient for convergence.  For all structures,  relaxations are preformed until the atomic Hellmann-Feynman forces are less than 1 meV/\AA \  and all in-plane stresses are less than 5 MPa.  

Table \ref{tab:table1} shows our results for four different
sheets:  the flat and buckled triangular sheets 
\cite{Evans05, Kunstmann06}, the hexagonal sheet, and one of our 
sheets ($\alpha$ in Figure \ref{fig:AB}).  The hexagonal
sheet is unstable with respect to in-plane shear, so we obtain the values 
in Table \ref{tab:table1} by  maintaining hexagonal symmetry while
optimizing the bond length. The binding energy is 
\[
E_b = E_{at} - E_{sheet},
\]
where $E_{at}$ is the energy of an isolated spin-polarized boron atom and
$E_{sheet}$ is the energy per atom of a sheet. The buckled triangular sheet is
more stable than the flat one due to the former forming
 stronger $\sigma$ bonds along the buckled direction 
\cite{Kunstmann06}. We also can reproduce previous
results on BNT made from triangular sheets
\cite{Kunstmann06,Evans05}.   

\begin{table}[t!]
\caption{\label{tab:table1} Binding energies $E_b$ (in eV/atom) and geometric 
parameters (in \AA) of four BS: the flat and buckled triangular
sheets, the hexagonal sheet, and one of our sheets ($\alpha$ in Figure
\ref{fig:AB}). $d^{flat}$ is the bond length of the flat triangular sheet.
$d^{\sigma}$ and $d^{diag}$ are the bond lengths of the buckled triangular
sheet, $d^{\sigma}$ is between atoms with the same z, while $d^{diag}$ is
between atoms with different z. $\Delta z$ is the buckling height. $d^{hex}$ is the bond length for the hexagonal sheet. $d^{new}$ gives the bond  
length range in the sheet $\alpha$.} 
\begin{tabular}{c l c c l c c c}
  \hline \hline 
 \  &    \multicolumn{2}{c}{Flat triangular} & \qquad & \multicolumn{4}{c}{Buckled triangular}\\ \cline{2-3} \cline{5-8}
 \   & $E_b$ & $d^{flat}$ & \qquad & $E_b$ & $d^{\sigma}$ & $d^{diag}$ & $\Delta z$ \\
\hline
LDA & 6.58 & 1.68  & \qquad & 6.74  & 1.59 & 1.80 & 0.81 \\
previous LDA\cite{Kunstmann06}
 & 6.76\footnote{Boron's atomic spin-polarization energy of 0.26
 eV/atom explains the $E_b$ differences between
 \cite{Kunstmann06} and our work or \cite{Evans05}.} 
& 1.69 & \qquad & 6.94$^a$ & 1.60 & 1.82 & 0.82 \\
previous LDA\cite{Evans05} & 6.53  &  -  & \qquad & 6.79 & - & - & - \\
GGA & 5.79 & 1.70  & \qquad & 6.00  & 1.60 & 1.86 & 0.88 \\
previous GGA\cite{Lau07} & 5.48\footnote{While the absolute $E_b$ from \cite{Lau07} do not match our GGA results, $E_b$ {\it differences} among the sheets match very well.}
  & 1.71  & \qquad & 5.70$^b$  & 1.61 & 1.89 & - \\
\hline \hline
 \  &    \multicolumn{2}{c}{Hexagonal} & \qquad & \multicolumn{4}{c}{Sheet $\alpha$}\\ \cline{2-3} \cline{5-8}
\    & $E_b$ & $d^{hex}$ & \qquad & \multicolumn{2}{c}{$E_b$} & \multicolumn{2}{c}{$d^{new}$}\\
\hline 
LDA & 5.82 & 1.65  & \qquad & \multicolumn{2}{c}{6.86}  & \multicolumn{2}{c}{1.64-1.67}\\
GGA & 5.25 & 1.67  & \qquad & \multicolumn{2}{c}{6.11}  &
 \multicolumn{2}{c}{1.66-1.69} \\
previous GGA\cite{Lau07} & 4.96$^b$ & 1.68  & \qquad & \multicolumn{2}{c}{-}  & \multicolumn{2}{c}{-} \\
\hline \hline
\end{tabular}
\end{table}

Figure \ref{fig:AB} shows two examples of our  BS, which are more stable
 than the buckled triangular sheet by 0.12 ($\alpha$) and 0.08 ($\beta$)
 eV/atom.
All our sheets are metallic, flat and  composed
of mixtures of hexagons and triangles. 
Sheet $\alpha$ is the most stable structure in our library.

Our sheets can be obtained by removing certain atoms from a flat
triangular sheet. Each removal produces a hexagonal hole, generating a
mixture of hexagons and triangles.
We define a ``hexagon hole density'' 
\[
\eta = \frac{\textrm{No. of hexagon holes }}
       {\textrm{No. of atoms in the original triangular sheet}}.
\]
The triangular sheet has $\eta$=0, the  hexagonal 
$\eta$=1/3, and sheets $\alpha$ and $\beta$ 
 have $\eta$ of $1/9$ and $1/7$, respectively.

\begin{figure}[b!]
\includegraphics[width=2.8in]{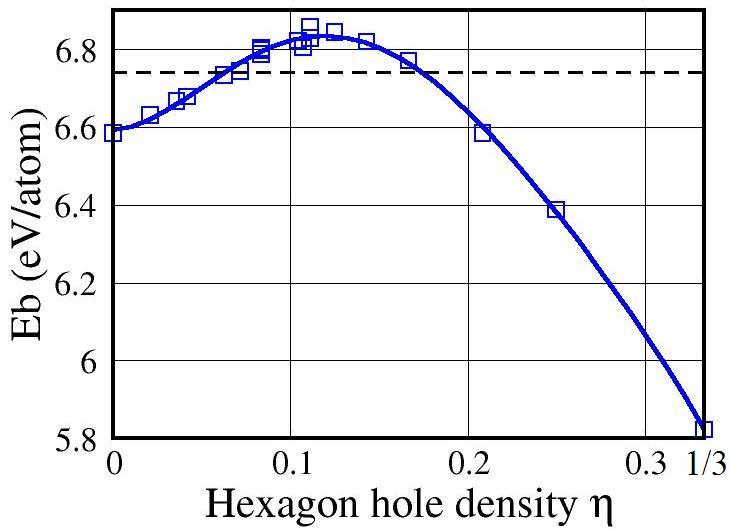}
\caption{\label{fig:doping}LDA $E_b$ v.s. hexagon hole
density $\eta$ for sheets with evenly distributed hexagons.  The  dashed line shows $E_b$ for the buckled triangular sheet. The solid blue curve is a polynomial fit. The two limiting cases $\eta=0$ and $\eta=1/3$ correspond to the flat triangular and hexagonal sheets, respectively.
Maximum $E_b$ occurs for sheet $\alpha$ ($\eta=1/9$).}
\end{figure}

{\it A priori}, the energies of these  sheets can depend on both $\eta$
and the pattern of hexagons. This results in a huge phase space of  hexagonal
patterns for a given $\eta$.  The most stable structures occur when the
hexagons are distributed as evenly as possible at fixed $\eta$.  Figure
\ref{fig:doping} shows the LDA binding energies $E_b$ versus $\eta$ for this
class of structures.  $E_b$ reaches a maximum of 6.86 eV/atom at $\eta$=1/9
(sheet $\alpha$). We also have investigated the other extreme where hexagons
form lines (\textit{e.g.}, sheet $\beta$). These ``linear'' structures are
more stable than the buckled triangular sheet for $\eta\approx$1/9 but are
less stable than the ``evenly-distributed'' class described above.  

\begin{figure}[b!]
\includegraphics[width=2.8in]{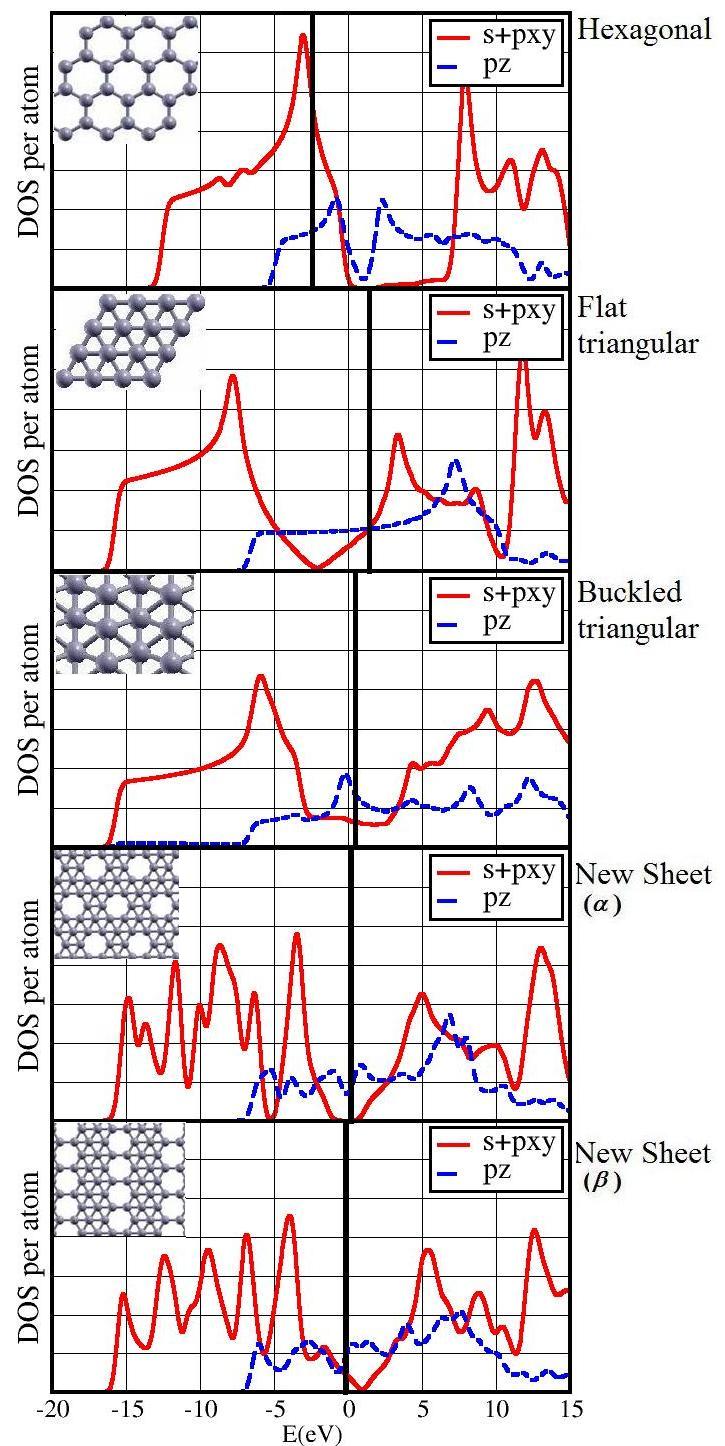}
\caption{\label{fig:ldos}Projected Densities of States (PDOS) for four BS. DOS
  are projected to in-plane (sum of $s$, $p_x$ and $p_y$) 
and out-of-plane orbitals ($p_z$). Red solid lines show in-plane and
blue dashed lines show out-of-plane projections; thick vertical solid
 lines show E$_F$. All curves are broadened
using Gaussians with a width of 0.3 eV. The vertical scale is
arbitrary. } 
\end{figure}

To explain the stability of these structures, we describe the nature
of their electronic bonding.  Generally,  in-plane bonds
formed from overlapping $sp^2$ hybrids are stronger than out-of-plane
$\pi$-bonds derived from $p_z$ orbitals, so a structure that
optimally fills in-plane bonding states should be most preferable.  Guided by
this principle, Figure \ref{fig:ldos} shows projected densities of
states (PDOS) for five BS with separate in-plane (the sum of
$s$, $p_x$ and $p_y$) and out-of-plane ($p_z$) projections.

We begin with the hexagonal sheet, a textbook $sp^2$ bonded system. 
All $sp^2$ hybrids are oriented
along nearest neighbor  vectors so that overlapping hybrids
produce canonical two-center bonds.  A large splitting ensues between 
in-plane bonding and anti-bonding states.
The $p_z$ orbitals form their own manifold of bonding and anti-bonding states.
The $p_z$ PDOS vanishes at the transition point between the two.
In the case of graphene, the four valence electrons per atom 
completely fill the $sp^2$ and the $p_z$ 
bonding states, 
leading to a highly stable structure. However, a boron atom has only three valence 
electrons. As shown in Figure \ref{fig:ldos}, some of the strong in-plane
$sp^2$ bonding states are 
unoccupied, explaining the instability of this sheet.  
For our discussion below, this sheet is highly prone to accepting electrons 
to increase its stability should they be available from another source. 

\begin{figure}[b!] 
\includegraphics[width=2.8in]{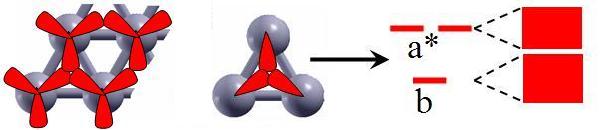}
\caption{\label{fig:flatsheet}
Three-center bonding scheme in flat triangular sheets.
Left: orientation of $sp^2$ hybrids. Center and right: overlapping
hybrids within a triangle ($D_3$ symmetry) yield one bonding ($b$) and two anti-bonding
($a^*$) orbitals. These then broaden into bands due to inter-triangle 
interactions.
}
\end{figure}

Next, we consider the flat triangular sheet.  Each atom has six nearest
neighbors but only three valence electrons. No two-center bonding
scheme leads to a proper description. Previous work has noted
qualitatively that a three-center bonding scheme exists here
\cite{Kunstmann06}.  We now present a detailed model of
the three-center bonding with crucial implications for the stability of our
sheets.  Figure \ref{fig:flatsheet} shows a choice of
orientations for the $sp^2$ hybrids where three hybrids overlap within an
equilateral triangle formed by three neighboring atoms.  
For an isolated such triangle, we have a simple 3$\times$3 
tight-binding problem with $D_3$ symmetry.  Its eigenstates are dictated by
group theory: one low-energy symmetric bonding orbital $b$ and two degenerate
high-energy anti-bonding orbitals $a^*$. (This is ``closed'' three-center bonding; details on this and other types of bonds are found in standard references \cite{Durrant62}.)  These orbitals then broaden into
bands due to inter-triangle couplings. 
Separately,  the $p_z$ orbitals also broaden into a single band (not shown).
In Figure \ref{fig:ldos}, the in-plane PDOS becomes zero at the energy
separating in-plane bonding and anti-bonding states. 
Ideally this sheet would be most stable if:  (i) two
electrons per atom would completely fill the $b$-derived in-plane bonding
bands, (ii) the anti-bonding $a^*$-derived bands were empty, and 
(iii) the remaining electron per atom would half fill the low-energy (bonding)
portion of the $p_z$-derived band. This would mean that the E$_F$ would be
at the zero point of the in-plane PDOS in Figure \ref{fig:ldos}.  Clearly,
this picture is a valid zeroth-order description.  However, E$_F$  lies
slightly above the ideal position and makes some electrons occupy in-plane
anti-bonding states. In other words, this sheet prefers to donate these
high-energy electrons which has critical implications below.  (Although we
seem to break symmetry by making half of the 
triangles filled and half empty, filling the entire $b$-derived in-plane
bonding band makes all hybrids  equally occupied.  This restores full 
in-plane symmetry:  {\it i.e.}, the two possible initial orientations of
hybrids give the same final state.)

The flat triangular sheet, however, buckles under small perturbations
along z \cite{Evans05}.   The buckling mixes in-plane and out-of-plane states and can be thought of as a symmetry reducing distortion that enhances binding.  As shown in Figure \ref{fig:ldos}, some states move below the
E$_F$ as indicated by the small peak immediately below the E$_F$.

Finally, we turn to the new structures.  The above discussion has shown that
the hexagonal sheet should be able to lower its energy by accepting electrons
while the flat triangular structure has a surplus of electrons in anti-bonding
states.   From a doping perspective, the three-center flat triangular regions
should act as donors while the two-center hexagonal regions act as acceptors.
Thus if the system is able to turn into a mixture of these two phases in the right
proportion, it should be able to benefit from the added stability of both
subsystems.  Specifically, the hexagon-triangle mixture with the highest
stability should be the one that places the E$_F$ precisely at the
zero-point of in-plane PDOS, filling all available in-plane
bonding states and none of the anti-bonding ones.  The remaining electrons
will fill the low-energy $p_z$-derived states, leading to a metallic system.
These expectations are born out clearly in Figure \ref{fig:ldos} as well as
in the energetic stability of the various structures (Figure
\ref{fig:doping}).  In fact, the most stable sheet, $\alpha$, satisfies this
condition precisely while the less stable sheet,
$\beta$, has a slight shift of E$_F$  from the ideal position.   

These findings have ramifications for boron clusters.  Our structures and bonding picture can explain that the extreme stability of B$_{80}$ fullerenes composed of triangular motifs with  pentagonal holes \cite{Szwacki07} is due to the well balance of three-center and two-center bonds.   Also the $\alpha$ sheet can be seen as the precursor of B$_{80}$ just as
graphene is the precursor of carbon fullerenes.  We also have studied some clusters. Figure \ref{fig:AB} shows the double-ring structures for B$_{24}$ and B$_{32}$ \cite{Boustani03, Boustani99} along with  clusters  constructed by us. The new B$_{24}$ cluster with a hexgon hole is less favorable by 0.08 eV/atom while the B$_{32}$ is {\em more} favorable by 0.03 eV/atom than the corresponding double-ring.  The stability of our sheets, of B$_{80}$, and our clusters with hexagonal holes suggests that for 
boron systems with more than 20-30 atoms, the Aufbau principle
breaks down and a more general structural rule is required.

In brief, we demonstrate a novel bonding mechanism in pure boron compounds arising from the competition between two- and three-center bonding.  This explains the stability of our boron sheets as well as larger boron clusters.  Our results have relevant implications on the stability and structure of boron clusters, boron nanotubes, and other boron systems.

We thank the Bulldog parallel clusters of the Yale High Performance Computing
center for providing computational resources.

\end{document}